\documentclass[aps,prd,onecolumn,superscriptaddress]{revtex4}
\usepackage{graphicx}
\usepackage{amsmath}
\usepackage{epsfig, epstopdf}
\usepackage{amsmath,amsfonts,amssymb}

\newcommand{\nc}{\newcommand}

\usepackage{color}
\nc{\ba}{\begin{eqnarray}}
\nc{\ea}{\end{eqnarray}}

\nc{\x}{{\bf{x}}}
\nc{\bfk}{{\bf{k}}}

\begin{document}
\baselineskip=12pt
\def\black{\textcolor{black}}
\def\red{\textcolor{black}}
\def\blue{\textcolor{blue}}
\def\green{\textcolor{black}}
\def\be{\begin{equation}}
\def\ee{\end{equation}}
\def\bea{\begin{eqnarray}}
\def\eea{\end{eqnarray}}
\def\orc{\Omega_{r_c}}
\def\om{\Omega_{\text{m}}}
\def\E{{\rm e}}
\def\bearst{\begin{eqnarray*}}
\def\eearst{\end{eqnarray*}}
\def\peleven{\parbox{11cm}}
\def\peffec{\peight{\bearst\eearst}\hfill\peleven}
\def\pspace{\peight{\bearst\eearst}\hfill}
\def\ptwelve{\parbox{12cm}}
\def\peight{\parbox{8mm}}

\title{Unraveling the nature of Gravity through our clumpy Universe}

\author{Shant Baghram}
\email{baghram-AT-sharif.edu}

\address{Department of Physics, Sharif University of
Technology, P.~O.~Box 11155-9161, Tehran, Iran}
\address{School of Astronomy, Institute for Research in
Fundamental Sciences (IPM),
P.~O.~Box 19395-5531,
Tehran, Iran}

\author{Saeed Tavasoli}
\email{tavasoli-AT-ipm.ir}
\address{School of Astronomy, Institute for Research in
Fundamental Sciences (IPM),
P.~O.~Box 19395-5531,
Tehran, Iran}

\author{Farhang Habibi}
\email{farhang-AT-ipm.ir}
\address{School of Astronomy, Institute for Research in
Fundamental Sciences (IPM),
P.~O.~Box 19395-5531,
Tehran, Iran}

\author{Roya Mohayaee}
\email{mohayaee-AT-iap.fr}
\address{IAP, CNRS, Sorbonne University, 98 bis Bld Arago, Paris, France}

\author{Joseph Silk}
\email{silk-AT-iap.fr}
\address{IAP, CNRS, Sorbonne University, 98 bis Bld Arago, Paris, France}

\vskip 1cm

\begin{abstract}
We propose a new probe to test the nature of gravity at various redshifts through large-scale cosmological observations.
We use our void catalog, extracted from the Sloan Digital Sky Survey (SDSS, DR10), to trace the distribution of matter along the lines of sight to SNe Ia that are selected from the Union 2 catalog. We study the relation between SNe Ia luminosities and  convergence and also the peculiar velocities of the sources.
We show that the effects, on SNe Ia luminosities, of convergence and of peculiar velocities predicted by the theory of general relativity and theories of modified gravities are different and hence provide a new probe of gravity at various redshifts.
We show that the present sparse large-scale data does not allow us to determine any statistically-significant
deviation from the theory of general relativity but future more comprehensive surveys should provide us with means for such an exploration.
\end{abstract}

\maketitle

\begin{center}

Essay written for the Gravity Research Foundation 2014 Awards for Essays on Gravitation\\
\end{center}


\section{INTRODUCTION: Cosmology as an arena for the study of the nature of gravity}

Gravity is the first force ever studied by physicists, but it is the last one
to be fully understood. The electromagnetic, weak and strong forces
are well formulated in the framework of quantum field theory, but gravity
and its classical description, the theory of general relativity (GR), are yet to be integrated into a unified picture of quantum physics.
It seems that quantum gravity or any other fundamental description of all the four forces
of nature is the holy grail of physics since the birth of general relativity and quantum mechanics in the 20th century.\\
Historically, the science of celestial mechanics and astronomy have enlightened
our understanding of gravity: the universal gravitational law
of Newton is based on celestial mechanics, and the perihelion precession of Mercury
has become a precise test of general relativity.

 Nowadays, the tradition works again.  Modern precision cosmology has provided us with
a huge opportunity to test and understand gravitational physics in the past three decades and it has also opened up new horizons to be explored.
The observation of the accelerated expansion of the Universe through
supernovae, as standard candles \cite{Perlmutter:1998np}, and the complementary observations
such as  the mapping of the cosmic microwave background radiation fluctuations (CMB)\cite{Ade:2013zuv} and large-scale structure (LSS)  \cite{Tegmark:2006az}
indicate that we are experiencing a kind of  ``antigravity'' in the Universe, which
could be caused by the cosmological constant (CC) or some mysterious mass-energy
in the Universe with negative pressure \cite{Peebles:2002gy}. However the discovery of the accelerated expansion of
the Universe could also indicate that the general theory of relativity
and the Einstein-Hilbert Lagrangian do not provide us with a correct classical theory of gravity,
and one needs to look further and search for modified theories of gravity.

The accelerated expansion of the Universe
along with its cosmological constant solution known as $\Lambda$CDM are based on  three assumptions.
Firstly, we have the assumption of the {\it{cosmological principle}} which
states that the Universe is homogeneous and isotropic on large scales.
Secondly, there is the assumption of   {\it{Einstein's theory of general relativity }} which is asserted to be
the correct theory of gravity in the classical limit. Thirdly, there  is the assumption that the
Universe is made up of components which interact through gravity, dark matter and baryons.
Any alternative to the cosmological constant can be categorized as a solution which breaks one
of the above assumptions. Non-homogeneous models \cite{Bolejko:2011jc}, dark energy theories \cite{Caldwell:1997ii}
and modified gravity (MG) models \cite{Clifton:2011jh}
provide alternative explanations for the accelerated expansion of the Universe. Cosmological observations which can distinguish between these three categories of models have been pursued
since the discovery of the accelerated expansion.\\
 There are two kinds of cosmological observations that can be used to distinguish between these models. First are the
geometrical observations which measure the distances in the Universe, for example
through supernovae Ia (SNe Ia) as standard candles
or statistically by measuring the abundance of structures to find the baryon acoustic oscillation (BAO) scale
as a standard ruler or the first peak of the CMB  \cite{Weinberg:2012es}. The second category of observations are the
dynamical probes which measure the growth of structure, such as
the power spectrum of the galaxies, the growth rate of the structures and  the cosmic shear of gravitational lensing. The dynamical observables probe the evolution of structures in different redshift ranges and on different scales (for a review of different cosmological models and their observational fingerprints in the light of the future Euclid mission, see \cite{Amendola:2012ys}). As the  growth of structures in the $\Lambda$CDM  is  scale-independent, any observation of scale-dependent growth can be a signature of deviations from standard gravity \cite{Baghram:2010mc,Mirzatuny:2013nqa}. \\

However the measurements of the matter power-spectrum $P_m(k,z)=|\delta^2_m(k)|$ or the growth rate $f\equiv d\ln\delta_m/d\ln a$, (where $a$ is the scale factor, $\delta_m=\rho/\bar{\rho}-1$ is the matter density contrast and $\bar{\rho}_m$ is the background density) are difficult and time-consuming tasks, where a considerable amount of statistics is required \cite{Tegmark:2006az} for the determination of the correlation function of galaxies and the redshift-space-distortion effect \cite{Kaiser:1987qv}.

  In this essay, we introduce a novel observational technique to measure
the growth of  structure and the matter power spectrum of galaxies on different scales.
We assume that the SNe Ia are standard candles and the background expansion of the Universe is given by the $\Lambda$CDM
model. This means that modified gravity theory must predict almost the same background expansion. Consequently, in order to distinguish between CC and MG, one should study the perturbations. In order to probe the evolution of the structures, we use the difference between the observed distance moduli of the SNe Ia and the theoretical predictions for the background expansion. This difference emerges from the convergence and/or  de-convergence of the light rays by the structures between the sources and the observers at high redshifts and also by the doppler lensing effect of peculiar velocities of the sources.\\ We show that at low redshifts the doppler lensing effect, due to peculiar velocities, is dominant  while at  intermediate redshifts, the two effects can be comparable. At lower redshifts, by measuring the peculiar velocities independently ({\it e.g} through the linear theory) we can estimate and compare the predictions of the standard model and alternative MG theory for the magnitude change in comparison to the background. At intermediate redshifts, the amount of lensing(de-lensing) of light bundles of standard candles can be extracted by knowing the amount of peculiar velocity corrections. Accordingly, we can find out about the evolution of structures over all redshift ranges from the observer to the source. In order to find the lensing/(de)lensing maps, we used a void catalog obtained by Tavasoli et al \cite{Tavasoli:2012ai} to find the voids and structures in Sloan Digital Sky Survey (SDSS) data release 10 (DR10) to measure the distribution of structures and the convergence observationally.( For more details see Tavasoli et al. (2014) \cite{Tavasoli2014} )\\
The structure of this essay is: In Sec.(\ref{Sec2}) we show how the perturbations affect the distance moduli of SNe Ia from its background evaluation and we will show how this will be related to the density contrast along the  line of sight.
In Sec.(\ref{Sec3}) we parameterize the deviation from Einstein GR via two parameters: the effective gravitational constant and the gravitational slip parameter. In Sec.(\ref{Sec4}), we discuss how the convergence-correction to luminosity distance can be used as a probe to study any deviation from GR. In Sec.(\ref{Sec5}) we show how we can find the convergence parameter observationally and finally in Sec.(\ref{Sec6}) we conclude and show that with future observations we should be able to study  the nature of gravity with the method introduced here.

\section{Convergence and the effect of peculiar velocities in a Clumpy Universe}
\label{Sec2}
As mentioned in the introduction, the cosmological observations
which are used to measure the background expansion of the Universe indicate that
$\Lambda$CDM is the best fit to the data at the background level. However, in order to probe the growth and evolution of structures in the Universe, we need to solve the Einstein equations with conservation laws at the perturbative level. This enables us to find the observational fingerprints for alternative models of CC
(such as dark energy/modified gravity). Furthermore, any deviation from GR by preserving the Lorentz Invariance of the theory will introduce a new degree of freedom \cite{Weinberg:1980kq}. This new degree of freedom by itself introduces a characteristic scale which determines the deviation of the rate of the growth of structures from that predicted by the $\Lambda$CDM. Consequently, it is worth studying the evolution of perturbations in order to distinguish between MG and CC.\\
On the other hand, observationally, we find from LSS surveys that the Universe is  almost but not exactly homogeneous and isotropic on large scales
($L> 100 Mpc$). The cosmological principle is an approximation because the structures in the Universe, such as  clusters of galaxies, group of galaxies and voids, make the cosmological principle  an approximation.
Consequently, we are obliged to study the clumpy Universe. Therefore, the expansion of the Universe and the evolution of the structures can be studied within the framework of a perturbed Friedmann-Robertson-Walker (FRW) metric:
\be
ds^2=-[1+2\Psi(\vec{x},t)]dt^2+a^2[1-2\Phi(\vec{x},t)]d^2\chi,
\ee
where $\Psi(\vec{x},t)$ and $\Phi(\vec{x},t)$ are the perturbed metric perturbations in the Newtonian gauge. In contrast to the background metric, they depend not only on the cosmic time, but also depend on position.
Now in order to study the evolution of perturbations, we use the perturbed Einstein equations $\delta G_{\mu\nu}=8\pi G \delta T_{\mu\nu}$, where we
can relate the metric perturbations $\Psi$ and $\Phi$ to the energy-momentum  of the constituents  of the Universe. Accordingly, the perturbed density contrast $\delta$ or the peculiar velocity of cosmic fluid act as the source of metric perturbations.
The Poisson equation (time-time and space-time components of Einstein field equations) relates the metric perturbation (gravitational potential $\Phi$ in the Newtonian gauge) to the gauge-invariant density contrast $\Delta_m$ in Fourier space as:
\be \label{eq:poiss}
k^2\Phi = 4\pi G \,a^2 \bar{\rho}_m\Delta_m,
\ee
where $k$ is the Fourier mode wavelength, the gauge invariant density contrast is $\Delta_m=\delta_m+ 3H\Theta (1+\omega)/k^2$, with equation of state $\omega\simeq 0$ for $\delta_m$ the density
contrast of a fluid (dark matter) and $\Theta$ is
the divergence of the peculiar velocity ${\bf{v}}$ of the fluid ({\it i.e.} $\Theta=\nabla . {\bf{v}}$).
Now any cosmological observation which is affected by the gravitational potential or by the density contrast of the matter in the Universe, can be used as a probe of the validity of the Poisson equation.
One of the main effects of a clumpy Universe is on the propagation of light.
The light bundles emitted from a cosmological source like a SN Ia undergo
two effects. First, the source is magnified or demagnified due to the over-dense (groups of galaxies)
and under-dense (voids) regions along the line of sight and the second effect is through the shear (the distortion of images).
These two effects can be quantified by a 2-dimensional mapping between
the flux of source $f^s$ and the flux reached to the observer $f^{obs}$ by a distortion matrix $A_{ij}$ as
$f^{obs}(\theta_i)=f^{s}(A_{ij}\theta_j)$
where $\theta_i$ is the angle which we observe the source, and $\theta_j$ in left hand side is the angle which shows the position of the source before lensing. The distortion matrix $A_{ij}$ is defined as:
\be
A_{ij}=\left(
         \begin{array}{cc}
           1-\kappa_g - \gamma_1 & \gamma_2 \\
           \gamma_2 & 1-\kappa_g + \gamma_1 \\
         \end{array}
       \right),
\ee
where $\kappa_g$ is the converge factor and $\gamma_{1,2}$ are the shear components.
The convergence is obtained from solving the spatial part of the geodesic equations. This is because we are interested in the light ray path in the clumpy universe. Using the definition of distortion matrix we can find the divergence as \cite{Dodelson:2003ft}
\be
\kappa_g = \frac{1}{2}\int_0^{\chi_s}d\chi (\chi_s - \chi)\frac{\chi}{\chi_s}\nabla_{\perp}^2 (2\Phi (\vec{x},t)),
\label{eq:kg}
\ee
where $\nabla_{\perp}$ is the two-dimensional derivative, which can be replaced by its 3D version in an approximation. The above equation is obtained by the assumption of GR, where the  non-diagonal equation $G_{\mu\nu}=8\pi G T_{\mu\nu}$ gives  $\Phi=\Psi$. We will relax this assumption in the next section and will probe the effect of this modification on convergence. Now by using Eq.(\ref{eq:kg}) and the Poisson Eq.(\ref{eq:poiss}) we can find  $\kappa_g$ along an arbitrary line of sight, by specifying the distribution of matter along the line of sight. This will open up a new horizon to study the distribution of matter at different redshifts. The recent large-scale structure surveys like SDSS\cite{Tegmark:2006az}, and future surveys like Euclid\cite{Amendola:2012ys}, and the Large Synoptic Survey Telescope (LSST)\cite{Abate:2012za}, will map the Universe on  large angular scales and at deep redshifts.

 The other important tracer of the matter distribution is the peculiar velocities of the SNe Ia host galaxies. The peculiar velocity can be obtained in linear perturbation theory by conservation of  energy:
\be
\delta'=-\vec{\nabla}.\vec{v}_p
\ee
where $'$ is the derivative with respect to conformal time and $\vec{v}_p$ is the peculiar velocity which can very roughly be approximated by
\be \label{pec-app}
{\vec{v}}_{p}\simeq Hr_s\delta(x)
\ee
where $r_s$ is the distance of the dark matter tracer ({\it i.e.} galaxy) from the center of the over/under
dense regions (in the linear regime), $H$ is the Hubble parameter at the redshift of the
source and $\delta(x)$ is the density contrast of the over/under dense region.\\
Eq.(\ref{pec-app}) is a very rough approximation of the peculiar velocity estimation from the Fourier transform of the continuity equation which is related to the growth rate of the structures $f\equiv d\ln\delta/d\ln a$ as below
\be \label{vpec}
{\vec{v}}_p(x)=\frac{iH}{(2\pi)^3}\int f(z)\delta_k\frac{\vec{k}}{k^2}e^{i\vec{k}.\vec{x}}d^3k
\ee
where the growth rate function in the dark-matter-dominated era is $f(z)=1$, and in the $\Lambda$CDM can be approximated as $f(z)=\Omega_m^\gamma(z)$, where $\gamma\simeq 0.55$. Accordingly, by knowing the growth of the structures and the local density contrast and the distance of the DM tracer from the over/under dense regions, we can estimate the peculiar velocities.

In the next Section, we introduce an almost general parametrization for MG theories, and their modification of the convergence parameter.

\section{Departure from the General theory of Relativity (GR)}
\label{Sec3}
In this section, we show how departures from  general relativity can be parameterized.
The first effect of modified gravity theories such as higher-dimensional models (like DGP \cite{Dvali:2000hr}),
$f(R)$ gravity theories \cite{DeFelice:2010aj}, massive gravity theories \cite{deRham:2014zqa} or the Galileon theories \cite{Nicolis:2008in}, is to modify the Poisson law. (For a review of MG models and their observational probes see \cite{Amendola:2012ys,Clifton:2011jh})
The modified Poisson Equation  in Fourier space can be parameterized as:
\be
k^2\Phi = 4\pi G a^2 Q(k,z)\bar{\rho}_m\delta_m,
\ee
where $Q(k,z)$ is the modification parameter measuring the departure from Einstein gravity, (As we study the sub-horizon scale, we can assume $\Delta_m\simeq \delta_m$). This parameter,  as we discussed in Sec.(\ref{Sec2}), introduces a scale-dependence, and that is why we introduce the modification parameter $Q(k,z)$ as a function of redshift and Fourier wave-mode $k$. In another way, to express this modification one can also
define an effective gravitational constant $G_{eff}=Q(k,z)G$, where in the case of $Q(k,z)=1$, Einstein gravity is recovered. The other modification of the governing equations arises from the non-diagonal space-space Einstein equations.
As we discussed earlier in the standard case, with assumption of a shear-less cosmic fluid we have $\Psi=\Phi$, where in the modified theories of gravity, we have a modified version of this relation as:
\be
\frac{\Phi}{\Psi}=\gamma(k,z)\,
\ee
where $\gamma (k,z)$ is the so-called gravitational slip parameter, which is scale-dependent like the effective gravitational constant.
Any deviation from GR in linear order perturbation theory is parameterized by  $\gamma (k,z)$ and $Q(k,z)$, and
consequently, it can be traced in the observations which deal
with the Poisson equation and dynamics of scalar metric perturbations \cite{Zhao:2008bn}.
Now the convergence as an observational probe of the distribution of matter along the line of sight, which has an effect on  light propagation, is modified. Equation (\ref{eq:kg}) can now be expressed as $
\kappa_g = {1}/{2}\int_0^{\chi_s}d\chi (\chi_s - \chi){\chi}/{\chi_s}\nabla_{\perp}^2(\Psi+\Phi)
$,
where $2\Phi$ is replaced by $\Phi+\Psi$.
The appearance of two gravitational potentials indicates the emergence of
the gravitational slip parameter $\gamma (k,z)$ and accordingly, the relation between $\Phi$ and  matter density
will  have the appearance of an effective Newtonian gravitational constant $G_{eff}=Q(k,z)G$.
The other effect of modified gravity theories is the change of the growth of  structures $f$. In the $\Lambda$CDM model, the growth rate is scale-independent. However in the MG theories, a characteristic scale is introduced in the modified Newtonian constant and gravitational slip parameter. Accordingly, the growth rate can be re-expressed by $f(k,z)=\Omega_m(z)^{\gamma_{MG}(k,z)}$, where $\gamma_{MG}(k,z)$ can be introduced for each model, and by inserting it into Eq.(\ref{vpec}), we can find the corresponding peculiar velocities of DM tracers.

In the next section we will show how we can use convergence along the line of sight to SNe Ia as a probe of modified gravity.


\section{The magnification change of SNe Ia as a test of gravity }
\label{Sec4}

In this section, we propose a new method to test  gravity at cosmological scales. We assume that
the background expansion of the Universe is well-described by the Hubble parameter obtained from the $\Lambda$CDM model.
Consequently, we can use the SNe Ia as standard candles to probe the deviations from $\Lambda$CDM prediction and interpret them as the effect of line-of-sight physics (like the gravitational lensing effect).
The distance moduli of any source is related to the luminosity distance $d_{L}$ through
\be
\mu=m-M=5\log_{10}d_{L}(z_s)+25 ,
\end{equation}
where $d_{L}$ is the luminosity distance in Mpc units and it is related to the comoving distance $\chi$ and angular diameter distance $d_A$ as $d_{L}=(1+z)\chi=(1+z)^2d_{A}$.
The luminosity distance of SNe Ia  can be written as $d_{L}=\bar{d}_{L}+\delta d_{L}$. The $\bar{d}_{L}$ is the luminosity distance in a homogeneous and isotropic universe. \\ $\delta d_{L}$ is introduced due to convergence
$\kappa_g$, (the effect of over-dense and under-dense regions on the propagation of light discussed in Sec. (\ref{Sec2}) for case of GR, and in Sec.(\ref{Sec3}) in the case of modified gravity),  doppler $\kappa_v$ (this is introduced due to the peculiar velocity of the source and observer), the Sachs-Wolfe (SW) effect $\kappa_{SW}$ and  the
Integrated Sachs-Wolfe (ISW) $\kappa_{ISW}$ (which are related to the difference in the amount of  gravitational  potential along the line of sight).  Consequently, the perturbation terms $\delta d_{L}$ are \cite{Bacon:2014uja}, {\it i.e.},
\be
\delta d_{L}=\bar{d}_{L}\left[-\kappa _g - \kappa_ v - \kappa _{SW}- \kappa _{ISW}\right],
\ee
where $\bar{d}_{L}(s)$, the luminosity distance of the background, is related to the cosmological parameters by $\bar{d}_{L}(s) = (1+z_s)H_0^{-1}\int_0^{z_s} dz/(\Omega_m(1+z)^3 + \Omega _{\Lambda})^{1/2}$
with the best-fit parameters of $\Lambda$CDM for the density parameter of matter $\Omega_m$ and cosmological constant $\Omega_{\Lambda}$ respectively. $\kappa_g$ is the convergence defined by Eq.(\ref{eq:kg}) and the doppler lensing is defined as:
\be
\kappa_v=\frac{1+z_s}{H\chi_s}{\bf{v}}_0.{\bf{n}}+(1-\frac{1+z_s}{H\chi_s}){\bf{v}}_s.{\bf{n}}.
\ee
where ${\bf{v}}_s$ is the peculiar velocity of the source.
We can neglect the $\kappa_{SW}=2\Phi_s+(1+z_s) (\Phi_o-\Phi_s) /(H\chi_s)$ and  $\kappa_{ISW}=2/\chi_s\int_0^{\chi_s}d\chi\Phi'+2[1-(1+z_s)/(H\chi_s)]\int_0^{\chi_s}d\chi\Phi'$. This is because both of these effects are related to the change of gravitational potential and its amplitude. In the late-time Universe $z < 2$, the gravitational potential is almost constant since the induced change from the CC-dominated era is small. To justify neglecting the Sachs-Wolfe and Integrated Sachs-Wolfe effects, we should compare the gravitational potential with the density contrast which is the source of the convergence and peculiar velocity terms. The Poisson Eq.(\ref{eq:poiss}) gives a rough estimate that is always $\Phi \simeq (\ell_{str}/\ell_{hor})^2 \delta _m$, where $\ell_{hor}$ is the size of the observable horizon, and $\ell_{str}$ is the size of the structure. In the same footing, the time derivative of the gravitational potential is small as well.  Another way to put this is to say that the gravitational potential always remains at the perturbation level $\Phi\le 1$, while the matter density can reach up to $10^5$ for structures. There are some super-voids of up to $\sim 200 Mpc$, where this approximation will become weaker, but as  we are probing the deviation of distance moduli at low redshifts, where the maximum size of any truly empty voids (rather than  that mainly due to the sparseness of the original galaxy catalog) are $\sim 30 Mpc$ in radius \cite{Tavasoli:2012ai}, this approximation should hold.
Consequently by neglecting the $\kappa_{SW}$ and $\kappa_{ISW}$ effects,
the deviation in luminosity distance $\Delta\mu \equiv \mu^{obs}-\mu^{\Lambda CDM}$,  which is the difference
between the standard cosmology distance moduli (the background) from the observational distance modulus, is sourced by convergence and peculiar velocity.
Now we can use the expected variation of magnitude $\Delta {\mu}$  within the
light cone as an indicator of the line of sight density contrast distribution and the peculiar velocity of the source.
For each SNe Ia, we can find the magnification change with respect to the homogenous and isotropic background case and plot it as a function of converge $\kappa_g$ (the amount of magnification/demagnification) and $\kappa_v$ as an indicator of peculiar velocity as below:
\be
\Delta\mu=\mu-\bar{\mu}=5\log(1-\kappa_g-\kappa_v)
\ee
where $\kappa_g$ can be obtained from Eq.(\ref{eq:kg}), where by replacing the $\nabla^2\Phi$ with density contrast $\delta$. In the next section, we will describe, how in real space we can find the $\kappa_g$ for each line of the sight along which  a SN Ia resides in SDSS space. The peculiar velocity can be found by different methods, which will be discussed in the next section. An independent way to measure the peculiar velocity is in linear perturbation theory by using Eq.(\ref{vpec}). Now it is obvious that by changing the gravity theory, both  $\kappa_g$ and $\kappa_v$ will be modified. In the calculation of $\kappa_g$  the modified Poisson equation is used and also the sum of the scalar perturbations appears in this calculation. Consequently, we can probe  deviations from the standard model via measuring the $Q$ and $\gamma$. In order to see the effect of this change, we can use the square root of the expected value of the $\kappa_g$ to show the modified gravity effect more profoundly\cite{Bacon:2014uja}.

\be \label{eq:kg2}
\kappa^2_g=\frac{9}{4}\Omega_m^0H_0^4\int_0^{\chi_s}d\chi
\left[(1+z)\frac{\chi_s-\chi}{\chi_s}\chi\right]^2\int_0^{\infty}\frac{k dk}{2\pi}P_m(k,z),
\ee
where $P_m(k,z)$ is  the {\it{effective}} matter power-spectrum.

 The matter power spectrum that appears in Eq.(\ref{eq:kg2}) can be modified due to modification of  gravity. In the most general case the matter power spectrum can be written in terms of the standard $\Lambda$CDM matter power spectrum ($P^{\Lambda CDM}_{m}$) as below:
\be
P_m(k,z)=\left[\frac{D^{MG}(k,z)}{D(z)}Q^2(k,z)(1+\gamma^{-1}(k,z))^2 \right]P^{\Lambda CDM}_m(k,z),
\ee
where $Q(k,z)$ is the ratio of the effective Newtonian constant to the bare gravitational constant; this term appears because we relate the gravitational potential to density contrast via the modified Poisson equation. Also
$\gamma(k,z)$ is the gravitational slip parameter, the term $(1+\gamma^{-1}(k,z))$ appears as the convergence is related to the integral of the two-dimensional divergence of $\Psi+\Phi$. $D(z)$ is the growth function, which shows how the density contrast grow from initial conditions. ( $\delta_m=D(z) \delta_m^{ini}$).
$D^{MG}(k,z)$ is the growth function in any chosen modified gravity. The important point here is that the growth function is a function of scale and redshift in contrast to the $\Lambda$CDM case where it is a scale-free function.\\
On the other hand, $\kappa_v$ can be used as a probe of  gravity as well. Eq.(\ref{vpec}) shows that the peculiar velocity is related to the growth rate function.
In the next Section, we will discuss how we can measure the three observational parameters $\Delta {\mu}$, $\kappa_g$ and $\kappa_v$. The knowledge of these terms through observations would allow us to compare
the predictions of a given model of modified gravity specifically by fixing the quantities
$D^{MG}(k,z)$ (interchangeably $f(k,z)$), $Q(k,z)$ , $\gamma(k,z)$. For example any deviation from
$Q(k,z)=\gamma(k,z)=1$ necessarily indicates a deviation from the theory of general relativity.


\section{Observational prospects}
\label{Sec5}

\begin{figure}
\includegraphics[width=10cm,natwidth=20cm,natheight=20cm]{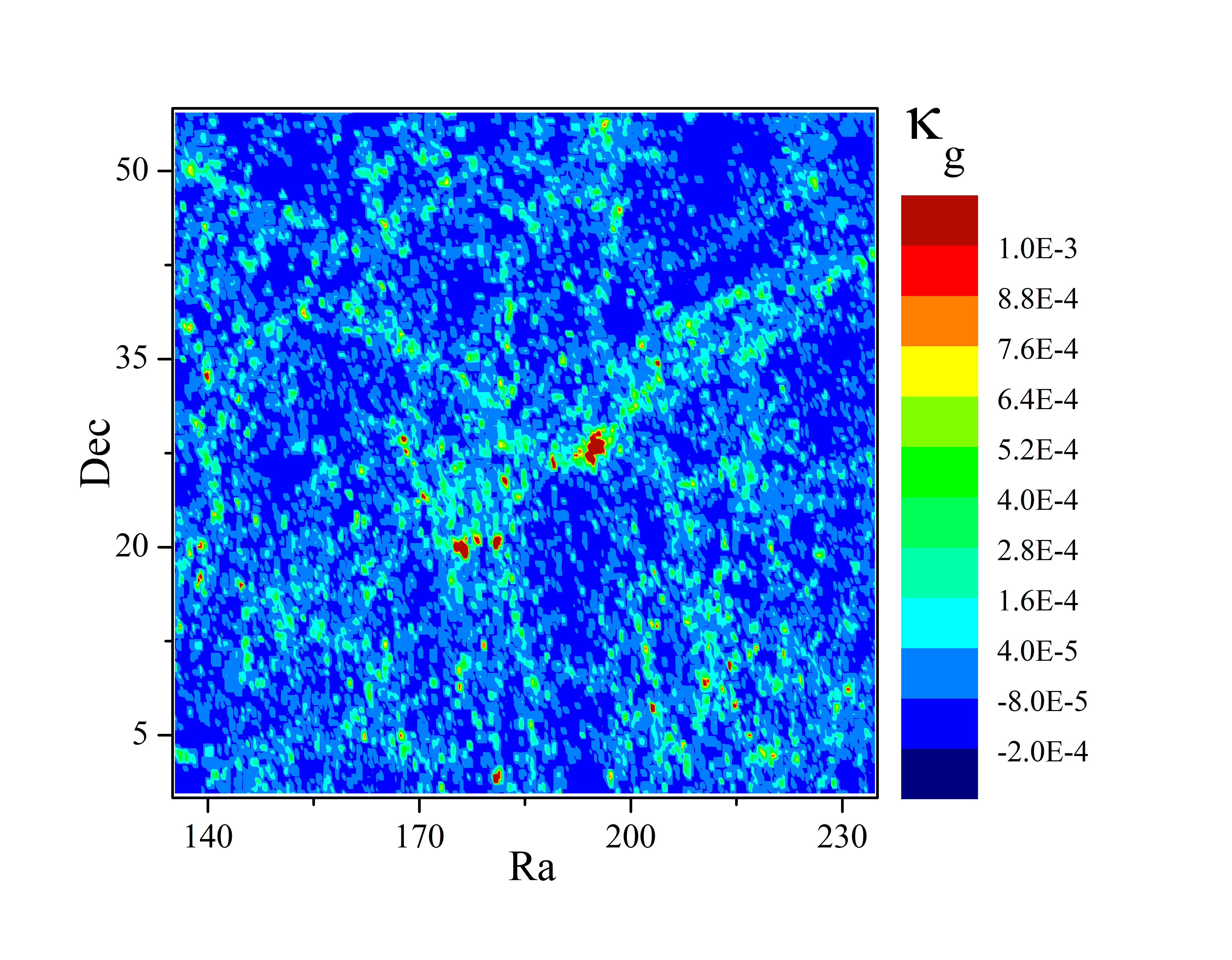}
\caption{In this figure we plot $\kappa_g$ for the $\Lambda$CDM model in the field of view of the SDSS integrated from redshift $0.01$ up to $0.04$. (The high value of $\kappa_g$ in declination of $\sim 27$ is related to the Coma cluster). $\kappa_g$ is in order of $10^{-3}$.}\label{figkg}
\end{figure}

\begin{figure}
\includegraphics[width=9cm,natwidth=20cm,natheight=20cm]{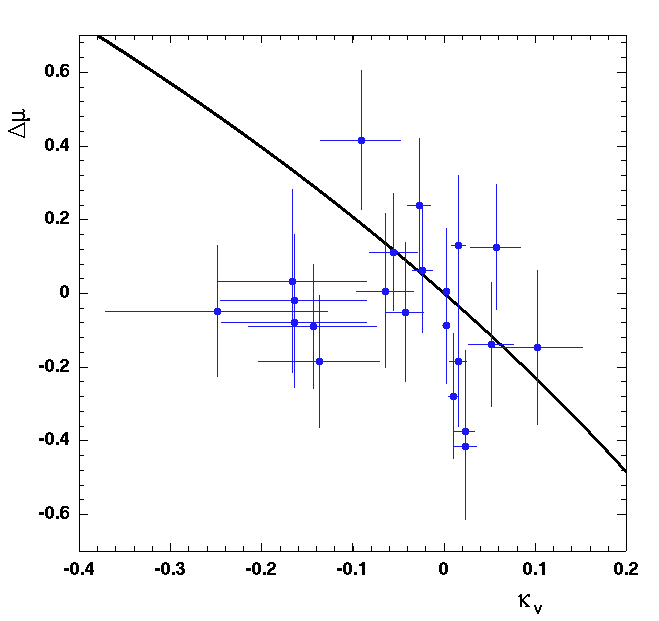}
\caption{In this figure we plot  $\Delta \mu$ for the $\Lambda$CDM model versus $\kappa$ where $\kappa=\kappa_v+\kappa_g$. $\kappa_v$ is obtained from Eq.(\ref{pec-app}) for 22 SNe Ia inside voids. $\kappa_g$ is calculated for $\Lambda$CDM by Eq.(\ref{eq:kg}) for the lines-of-sight to the  SNe Ia, and is 2 orders of magnitude smaller than $\kappa_v$ and can be neglected. The error bars on $\Delta\mu$ is obtained from the Union sample. The error bars on $\kappa_v$ come from the distance errors of the SNIa host galaxies from the center of the under-dense region. (The error bars are $1\sigma$).}\label{figkappazd}
\end{figure}

In order to find the values of $\Delta\mu$ we use the Union2
catalog \cite{Amanullah:2010vv} which contains 557 SNe Ia of which
192 are at $z < 0.15$ (the highest redshift that we probe). This is one of the largest compilations of SNe Ia
to date from several different surveys. We can find the $\Delta \mu$ for each SN Ia, using the concordance model, with the best parameter fit obtained from cosmological observations \cite{Hinshaw:2012aka}. The data are corrected for our peculiar motion in the CMB rest frame. Consequently from velocity corrections, only the peculiar velocity of the source will have a contribution in our analysis.
Now the crucial part of our work is the procedure that we use to extract $\kappa_g$.
In order to find the $\kappa_g$ along each line of sight to the SNe Ia we use the void catalog introduced by Tavasoli et al. \cite{Tavasoli:2012ai}. This catalog is based on the original algorithm of the Aikio
and Maehoenen (AM) \cite{Aikio1998} method in the 3D  sample. Using this method which is not biased to a specific shape for void finding, on the SDSS DR10 volume-limited galaxy sample \cite{Ahn:2013gms}, we obtain a void catalog up to redshift $z=0.15$. The selected region in the SDSS observational area is within a declination of $(0,55)$ and right ascension $(135,235)$.
To ensure statistical uniformity, we divide the sample in 4-sub samples with redshift ranges of $z=0.01-0.04 , 0.04-0.065, 0.065 - 0.10, 0.10 - 0.15$ and with maximum absolute magnitude value in the r-band $M_r= -18.5, -19.5, -20.5, -21.5$ correspondingly. In order to probe the SDSS catalog to higher redshifts that of $z>0.15$, we should choose galaxies brighter than $M<-22$ and consequently there will be a bias to very luminous galaxies which in turn gives very large voids. However future deeper surveys may provide the opportunity to have void catalogs at higher redshifts.
Now by using a grid of $1 Mpc h^{-1}$ resolution on the SDSS DR10 void catalog, we can trace the path of light rays from the source SNe Ia to us. The number of SNe Ia up to $z<0.15$ in selected region of ours in SDSS observational area is 35. The light rays from  these 35 SNe Ia  pass through each grid via an over-dense, or under-dense region. Having the void structure in the 3D observational sample gives us the possibility to measure the under-dense density contrast. (This is done by counting the field galaxies that are inside the void, and dividing by the void volume). Accordingly, by integrating the line of sight value of $\kappa_g ={3}/{2}\Omega_m^0H_0^2\int_0^{\chi_s}d\chi
\left[(1+z)\frac{\chi_s-\chi}{\chi_s}\chi\right]\delta_{+,-}/b$  we can find the value of  $\kappa_g$ observationally (We use the Poisson equation to relate the gravitational potential in Eq.(\ref{eq:kg}) to the density contrast along the  line-of-sight). \\$\delta_{+}$, $\delta_{-}$ relates to the over(under) dense regions respectively. We also add a bias parameter which is defined as $b\equiv \delta_m / \delta _{+,-}$. The bias parameter appears here as we find the voids with luminous matter (galaxies) and also we find the over-densities by galaxies instead of dark matter halos. The bias parameter can introduce more complications into the study of the cosmological models. It could be scale-dependent, redshift-dependent and also  an environment-dependent quantity \cite{Dalal:2007cu}. However we assume that in this work the bias is a constant value. \\
Consequently we can find the convergence along each line of sight to the SNe Ia.
In Fig. (\ref{figkg}) we plot the values of $\kappa_g$ for $\Lambda$CDM model
assuming $b=1$ in the redshift range of $0.01-0.04$ which is the first sub-volume limited sample of SDSS with galaxies $M_r=-18.5$ as the characteristic absolute value of galaxies. The Fig. (\ref{figkg}) is produced by Eq.(\ref{eq:kg}) as described in the above paragraph. By knowing the modification to the Eq.(\ref{eq:kg}) for any desired model of modified gravity, this figure can be reproduced. It is worth  mentioning that at low redshifts, $\kappa_g$ is on the order of $10^{-3}$, on average two orders of magnitude smaller than the peculiar velocity effect. The advantage of this method is that we can make $\kappa_g$ maps for any desired model for each observational LSS survey like the SDSS. \\
The other important observational quantity that we should find  is the peculiar velocity of the source. \\ The peculiar velocity of the source can be found by different methods such as:  a: distance measurement indicators like the luminosities of a particular class of galaxies. b:  redshift space distortions, c: the linear theory predictions\cite{Hudson:2012gt}. \\
 The peculiar velocities can be obtained from the distance measurements as:
\be \label{vpec1}
v_{pec}\simeq v_{obs}-\frac{H d_L}{(1+\bar{z})}
\ee
where $v_{obs}$ is the observed velocity and $d_{L}$ is the luminosity distance.
The important point to indicate here is that the distance measurement method is not a suitable method to test gravity. This is because, at low redshift, the gravitational convergence correction to the magnitude change is small and negligible. In the other hand in the distant measurement method, all the magnitude change from the background is assigned to the effect of peculiar velocities, consequently this gives a biased result as the measurement of $\kappa_v$ and $\Delta\mu$ will not be independent. Accordingly, we need  an independent way of measuring peculiar velocities.

The linear theory prediction which is itself modified due to deviation from GR can be obtained as:
\be \label{eq:pec}
\vec{v}(r) = \frac{iH_0}{(2\pi)^3}\int_0^{\infty} \frac{f^{MG}(k)}{b}\frac{\vec{k}}{k^2}D^{MG}(k,z)\delta_{m}(k,z=0)e^{i{\vec{k}}.{\vec{r}}}dk
\ee
where $D^{MG}(k,z)$ is the growth function $\delta_{m}(k,z=0)$ is the density contrast of matter at the present time
and $f^{MG}\equiv d\ln D^{MG}(k,z)/ d\ln a$.

For this work, we just use Eq.(\ref{pec-app}) to obtain the peculiar velocities and correspondingly the value of $\kappa_v$  for the SNe Ia. As the linear theory prediction for the peculiar velocity is an approximation for the regions with density contrast on order of unity or smaller, we choose a sample of  22 SNe Ia that are in voids. This is because the voids are in the linear regime to a  good approximation.
In Fig.(\ref{figkappazd}), we plot the distance modulus difference $\Delta \mu$ for 22 SNe Ia which are inside SDSS regions, and  reside inside voids, versus $\kappa$, where $\kappa=\kappa_g+\kappa_v$. As was mentioned before, $\kappa_g$ is obtained from the line-of-sight integration of density contrast and $\kappa_v$ is obtained from the linear theory approximation.




Fig. (\ref{figkappazd}) shows that the current data is not precise enough to show any statistical viable tension for the $\Lambda$CDM prediction for the $\Delta\mu - \kappa$ relation.
In this work we have just examined the standard model prediction, to show the procedure and the method in order to use $\Delta\mu - \kappa$ relation for testing the models. It is worth  mentioning that at low redshifts $z<0.2$,  $\kappa_v$ is the dominant effect, while at intermediate redshifts, both the effects of convergence/deconvergence become important.
Finally, it is obvious that probing   alternatives to standard gravity with this method is not efficient now due to the  limited coverage of the sky in the relevant redshift range andthe incompleteness of the  low redshift void-cluster catalogs.

However  future LSS surveys should provide us with the opportunity to determine  the $\kappa_g$ map of the Universe on larger angular scales and deeper in redshift. On the other hand, future SNe Ia hunters such as LSST, DES, etc. will increase the number of SNe Ia along the line-of-sight in any desired $\kappa_g$ map.
The mapping of the Universe with future LSS surveys will also let us  make convergence maps on larger scales and deeper in redshifts. Accordingly, this method can be used as a promising probe for tomography of the density contrast in the Universe and consequently as a promising observational technique for distinguishing between models of an accelerated expansion Universe and  tests of gravity.


\section{Conclusion and future prospects}
\label{Sec6}
In this essay, we propose a new method to test gravity on cosmological scales. SNe Ia as
standard candles have been used to probe distances and measure the rate of
the expansion of the Universe. As the background expansion of the Universe is
well-described by $\Lambda$CDM, the distinction between cosmological constant and modified theories of gravity
can only be probed at the perturbative level. The difference between the
observed distance modulii of SNe Ia and those given by the predictions of homogenous and isotropic $\Lambda$CDM is due to
perturbations along the  line-of-sight.
There are two important major sources of perturbations. First there is the peculiar velocity of
the source and the second is the convergence due to
the lensing(anti-lensing) effect of over(under) densities along the
line-of-sight.  The convergence is a promising quantity for distinguishing
between the theory of general relativity and modified gravity theories. This is because the
convergence depends on the metric perturbations $\Phi$ and $\Psi$. Consequently, we show that we can
check the Einstein equations (Poisson (time-time),(time-space)) and the non-diagonal component
(space-space) of $G_{\mu\nu}=8\pi GT_{\mu\nu}$ by this method.

In order to parameterize the theories of modified gravity, we have defined an effective gravitational constant
$G_{eff}$ and a gravitational slip parameter $\gamma$ and show how the convergence can be defined
in modified theories of gravity.
Observationally, we compute the quantity $\kappa_g$ using our
void-finding algorithm on the SDSS DR10 redshift catalog to find the over-dense
and under-dense regions along the line of sight. The volume-limited region of SDSS with 4 sub-volumes and the Union 2 sample of SNe Ia allow us to test our proposal with 35 SNe Ia.
On the other hand, we have argued that peculiar velocities have the dominant effect on the magnitude change of the SNe Ia at low redshifts. However, an independent measurement of the peculiar velocities is essential in order to measure the deviations from the background prediction through SNe Ia luminosities. Accordingly, we propose the use of linear perturbation theory for obtaining $\kappa_v$. In order to have a more viable approximation, we just use the SNe Ia that reside inside voids (The number of SNe Ia  spanned by SDSS which reside inside voids becomes 22). This is because the voids are almost in the linear regime.
The current data plotted in Fig.(\ref{figkappazd}), shows no significant deviation from $\Lambda$CDM,  perhaps because the statistics are very low.
 %
 Future observations will provide better statistics for a more precise determination of $\kappa_g$ as we can probe to higher redshifts. This will be possible because of the increase in the statistics and also via probing the Universe at higher redshifts where $\kappa_g$ becomes more important. On the other hand, more sophisticated methods to measure the peculiar velocity ({\it e.g.} \cite{mohetal}) can also be useful to test  gravity at low redshifts.

A combined study of the correlation between
large-scale structure and supernovae will enable us to probe the evolution of structures and
study the scale-dependence of growth of structures, and eventually to search for any deviations from the theory of general relativity.

\acknowledgments
We thank Hadi Rahmani, Ehsan Kourkchi,
Jean Philip Uzan , Brent Tully, Rahman Amanullah, Pierre Fleury
and Sohrab Rahvar for help and discussions.
This research has been supported in part by the Balzan foundation via visit of SB to Institut d'astrophysique de Paris (IAP).
We also thank
the anonymous referee for a careful reading of the manuscript and comments.


\begin{thebibliography} {50}

\bibitem{Perlmutter:1998np}
  S.~Perlmutter {\it et al.}  [Supernova Cosmology Project Collaboration],
  ``Measurements of Omega and Lambda from 42 high redshift supernovae,''
  Astrophys.\ J.\  {\bf 517}, 565 (1999)
  [astro-ph/9812133].

\bibitem{Ade:2013zuv}
  P.~A.~R.~Ade {\it et al.}  [Planck Collaboration],
  ``Planck 2013 results. XVI. Cosmological parameters,''
  arXiv:1303.5076 [astro-ph.CO].

\bibitem{Tegmark:2006az}
  M.~Tegmark {\it et al.}  [SDSS Collaboration],
  ``Cosmological Constraints from the SDSS Luminous Red Galaxies,''
  Phys.\ Rev.\ D {\bf 74}, 123507 (2006)
  [astro-ph/0608632].

\bibitem{Peebles:2002gy}
  P.~J.~E.~Peebles and B.~Ratra,
  ``The Cosmological constant and dark energy,''
  Rev.\ Mod.\ Phys.\  {\bf 75}, 559 (2003)
  [astro-ph/0207347].

\bibitem{Bolejko:2011jc}
  K.~Bolejko, M.~-N.~Celerier and A.~Krasinski,
  ``Inhomogeneous cosmological models: Exact solutions and their applications,''
  Class.\ Quant.\ Grav.\  {\bf 28}, 164002 (2011)
  [arXiv:1102.1449 [astro-ph.CO]].


\bibitem{Caldwell:1997ii}
  R.~R.~Caldwell, R.~Dave and P.~J.~Steinhardt,
  ``Cosmological imprint of an energy component with general equation of state,''
  Phys.\ Rev.\ Lett.\  {\bf 80}, 1582 (1998)
  [astro-ph/9708069].

\bibitem{Clifton:2011jh}
  T.~Clifton, P.~G.~Ferreira, A.~Padilla and C.~Skordis,
  ``Modified Gravity and Cosmology,''
  Phys.\ Rept.\  {\bf 513}, 1 (2012)
  [arXiv:1106.2476 [astro-ph.CO]].

\bibitem{Weinberg:2012es}
  D.~H.~Weinberg, M.~J.~Mortonson, D.~J.~Eisenstein, C.~Hirata, A.~G.~Riess and E.~Rozo,
  ``Observational Probes of Cosmic Acceleration,''
  Phys.\ Rept.\  {\bf 530}, 87 (2013)
  [arXiv:1201.2434 [astro-ph.CO]].




\bibitem{Amendola:2012ys}
  L.~Amendola {\it et al.}  [Euclid Theory Working Group Collaboration],
  ``Cosmology and fundamental physics with the Euclid satellite,''
  Living Rev.\ Rel.\  {\bf 16}, 6 (2013)
  [arXiv:1206.1225 [astro-ph.CO]].


\bibitem{Baghram:2010mc}
  S.~Baghram and S.~Rahvar,
  ``Structure formation in $f(R)$ gravity: A distinguishing probe between the dark energy and modified gravity,''
  JCAP {\bf 1012}, 008 (2010)
  [arXiv:1004.3360 [astro-ph.CO]].



\bibitem{Mirzatuny:2013nqa}
  N.~Mirzatuny, S.~Khosravi, S.~Baghram and H.~Moshafi,
  ``Simultaneous effect of  modified gravity  and  primordial non-Gaussianity in large scale structure observations,''
  JCAP {\bf 1401}, 019 (2014)
  [arXiv:1308.2874 [astro-ph.CO]].



\bibitem{Kaiser:1987qv}
  N.~Kaiser,
  ``Clustering in real space and in redshift space,''
  Mon.\ Not.\ Roy.\ Astron.\ Soc.\  {\bf 227}, 1 (1987).


\bibitem{Tavasoli:2012ai}
  S.~Tavasoli, K.~Vasei and R.~Mohayaee,
  ``The challenge of large and empty voids in SDSS DR7 redshift survey,''
  arXiv:1210.2432 [astro-ph.CO].



  \bibitem{Tavasoli2014}
 S.~Tavasoli, F.~Habibi, S.~Baghram, R.~Mohayaee and J.~Silk,
 {\it{In preperation}}



\bibitem{Weinberg:1980kq}
  S.~Weinberg and E.~Witten,
  Phys.\ Lett.\ B {\bf 96}, 59 (1980).



\bibitem{Dodelson:2003ft}
  S.~Dodelson,
  Amsterdam, Netherlands: Academic Pr. (2003) 440 p

\bibitem{Abate:2012za}
  A.~Abate {\it et al.}  [LSST Dark Energy Science Collaboration],
  ``Large Synoptic Survey Telescope: Dark Energy Science Collaboration,''
  arXiv:1211.0310 [astro-ph.CO].



\bibitem{Dvali:2000hr}
  G.~R.~Dvali, G.~Gabadadze and M.~Porrati,
  ``4-D gravity on a brane in 5-D Minkowski space,''
  Phys.\ Lett.\ B {\bf 485}, 208 (2000)
  [hep-th/0005016].

\bibitem{DeFelice:2010aj}
  A.~De Felice and S.~Tsujikawa,
  ``f(R) theories,''
  Living Rev.\ Rel.\  {\bf 13}, 3 (2010)
  [arXiv:1002.4928 [gr-qc]].

\bibitem{deRham:2014zqa}
  C.~de Rham,
  ``Massive Gravity,''
  Living Rev.\ Rel.\  {\bf 17}, 7 (2014)
  [arXiv:1401.4173 [hep-th]].


\bibitem{Nicolis:2008in}
  A.~Nicolis, R.~Rattazzi and E.~Trincherini,
  ``The Galileon as a local modification of gravity,''
  Phys.\ Rev.\ D {\bf 79}, 064036 (2009)
  [arXiv:0811.2197 [hep-th]].


\bibitem{Zhao:2008bn}
  G.~-B.~Zhao, L.~Pogosian, A.~Silvestri and J.~Zylberberg,
  ``Searching for modified growth patterns with tomographic surveys,''
  Phys.\ Rev.\ D {\bf 79}, 083513 (2009)
  [arXiv:0809.3791 [astro-ph]].


\bibitem{Bacon:2014uja}
  D.~J.~Bacon, S.~Andrianomena, C.~Clarkson, K.~Bolejko and R.~Maartens,
  ``Cosmology with Doppler Lensing,''
  arXiv:1401.3694 [astro-ph.CO].

\bibitem{Amanullah:2010vv}
  R.~Amanullah, C.~Lidman, D.~Rubin, G.~Aldering, P.~Astier, K.~Barbary, M.~S.~Burns and A.~Conley {\it et al.},
  ``Spectra and Light Curves of Six Type Ia Supernovae at 0.511 < z < 1.12 and the Union2 Compilation,''
  Astrophys.\ J.\  {\bf 716}, 712 (2010)
  [arXiv:1004.1711 [astro-ph.CO]].


\bibitem{Hinshaw:2012aka}
  G.~Hinshaw {\it et al.}  [WMAP Collaboration],
  ``Nine-Year Wilkinson Microwave Anisotropy Probe (WMAP) Observations: Cosmological Parameter Results,''
  Astrophys.\ J.\ Suppl.\  {\bf 208}, 19 (2013)
  [arXiv:1212.5226 [astro-ph.CO]].
\bibitem{Aikio1998}
Aikio, J. and Maehoenen, P. 1998, ApJ, 497, 534



\bibitem{Ahn:2013gms}
  C.~P.~Ahn {\it et al.}  [SDSS Collaboration],
  ``The Tenth Data Release of the Sloan Digital Sky Survey: First Spectroscopic Data from the SDSS-III Apache Point Observatory Galactic Evolution Experiment,''
  Astrophys.\ J.\ Suppl.\  {\bf 211}, 17 (2014)
  [arXiv:1307.7735 [astro-ph.IM]].

\bibitem{Dalal:2007cu}
  N.~Dalal, O.~Dore, D.~Huterer and A.~Shirokov,
  ``The imprints of primordial non-gaussianities on large-scale structure: scale dependent bias and abundance of virialized objects,''
  Phys.\ Rev.\ D {\bf 77}, 123514 (2008)
  [arXiv:0710.4560 [astro-ph]].


\bibitem{Hudson:2012gt}
  M.~J.~Hudson and S.~J.~Turnbull,
  ``The growth rate of cosmic structure from peculiar velocities at low and high redshifts,''
  Astrophys.\ J.\  {\bf 751}, L30 (2013)
  [arXiv:1203.4814 [astro-ph.CO]].

\bibitem{mohetal} Mohayaee, R., Mathis,
H., Colombi, S., and Silk, J., Mon.\ Not.\ Roy.\ Astron.\ Soc.\ , 365, 939, (2006)


\end{thebibliography}
\end{document}